# Enhanced Charge Transport in A-site Ordered Perovskite Derivatives $A_2A'Bi_2I_9$ (A = Cs; A'= Ag, Cu): A First-Principles Study

Shuhan Li[1,2], Siyu Song[1,2], Peng Lv[3*], Shihao Wang[1,2], Jiawang Hong[4], Gang Tang[1,2*]

[1]Advanced Research Institute of Multidisciplinary Science, Beijing Institute of Technology, Beijing 100081, China

[2]Beijing Institute of Technology, Zhuhai Beijing Institute of Technology (BIT) Zhuhai 519088, P.R.China

[3] Key Laboratory for High Efficiency Energy Conversion Science and Technology of Henan Province, International Joint Research Laboratory of New Energy Materials and Devices of Henan Province, School of Physics and Electronics, Henan University, Kaifeng, 475004, China

[4]School of Aerospace Engineering, Beijing Institute of Technology, Beijing, 100081, China

**\*E-mails**:
Peng Lv: lvpeng@henu.edu.cn;
Gang Tang: gtang@bit.edu.cn.

# Abstract

Recent experiments have synthesized $Cs_2AgBi_2I_9$ by partially substituting $Cs^+$ with $Ag^+$ at the A-site of $Cs_3Bi_2I_9$, resulting in enhanced charge transport properties compared to $Cs_3Bi_2I_9$. However, the atomic-scale mechanisms behind this enhancement remain unclear. In this work, we investigate the carrier transport mechanisms in $Cs_2A'Bi_2I_9$ (A' = Ag, Cu) using first-principles calculations and Boltzmann transport calculations. Our results reveal that A-site ordered $Cs_2A'Bi_2I_9$ exhibits carrier mobilities that are 3-4 times higher than those of $Cs_3Bi_2I_9$ within the 100-500 K temperature range. We identify polar phonon scattering as the dominant mechanism limiting mobility. Furthermore, the enhanced out-of-plane carrier mobility in $Cs_2A'Bi_2I_9$, particularly between 100 and 200 K, leads to reduced mobility anisotropy. These improvements are mainly due to the shorter A'-I bond lengths and increased $Ag^+/Cu^+$ s-I p orbital coupling. Notably, substitution with $Cu^+$ results in a further reduction in the band gap and enhanced hole mobility compared to $Ag^+$ substitution in $Cs_3Bi_2I_9$. Further analysis reveals that the significant increase in carrier mobility in $Cs_2A'Bi_2I_9$ can be largely explained by the smaller carrier effective masses ($m^*$) and weaker Fröhlich coupling strengths ($α$), resulting in a lower polar mass $α(m^*/m_e)$, compared to $Cs_3Bi_2I_9$. Our study provides valuable insights into the transport properties of Bi-based perovskite derivatives, paving the way for their future applications in optoelectronic devices.

Although lead-based halide perovskites have achieved numerous breakthroughs in optoelectronic applications, their commercialization is still hindered by two major factors: the presence of toxic lead and their long-term stability.[1,2] To address these challenges, researchers worldwide have been actively developing high-performance, highly stable, lead-free perovskite alternatives in recent years.[3-5] Among low-toxicity, perovskite-inspired materials, replacing Pb(II) with Bi(III) is of particular interest due to its environmentally friendly nature, $ns^2$ valence electron configuration, and intrinsic environmental stability.[6,7]

To maintain charge neutrality, Bi(III)-based perovskite variants usually adopt the chemical formula $A_3B_2X_9$, derived from the parent $ABX_3$ structure by removing 1/3 of the B-site cations. The $A_3B_2X_9$ compounds have two typical polymorphs: one is the hexagonal (H) phase with a space group of $P6_3/mmc$ that consists of face-sharing octahedral 0D dimers, and the other is the trigonal (T) phase with a space group of $P$-$3m1$ composed of a corner-sharing octahedral 2D layered structure.[8] Generally, the H phase is characterized by a wide band gap, high exciton binding energy, and large carrier effective masses.[9-12] These properties arise from the face-sharing octahedra, which weaken the $s$-$p$ orbital hybridization between the metal cation and halogen anion.[13]

Many studies have aimed to enhance the optoelectronic properties of the H phase by modifying the A-site cations, B-site cations, and X-site anions in 0D $A_3B_2X_9$ to increase structural dimensionality.[14-17] For example, in 0D $Cs_3Bi_2I_9$, introducing a monovalent $Ag^+$ cation at the B-site transforms it into the 3D $Cs_2AgBiBr_6$.[18] Additionally, forming anion ordering at the X-site of 0D $Cs_3Bi_2I_9$ to create the 2D $Cs_3Bi_2I_6Br_3$ results in enhanced carrier transport ability.[19,20] Very recently, Hossain et al. successfully synthesized $Cs_2AgBi_2I_9$ single crystals by partially substituting the A-site $Cs^+$ cation in $Cs_3Bi_2I_9$ with $Ag^+$. Compared to 0D $Cs_3Bi_2I_9$, the A-site ordered $Cs_2AgBi_2I_9$ exhibits a reduced band gap by 0.27 eV, weaker electron-phonon coupling, and a higher carrier mobility-lifetime product.[21] However, the dominant scattering mechanism limiting carrier mobility in the newly synthesized system, as well as the mechanisms responsible for its enhanced charge transport compared to $Cs_3Bi_2I_9$, remain unclear.

In this work, we investigate the carrier transport properties of A-site ordered $Cs_2A'Bi_2I_9$ (A' = Ag, Cu) using first-principles calculations combined with Boltzmann transport theory. Our calculations reveal that the introduction of $Ag^+$ or $Cu^+$ cations at the A-site of $Cs_3Bi_2I_9$ significantly weakens chemical bonds, reduces the band gap, and enhances band edge dispersion. Additionally, the carrier mobilities of A-site ordered $Cs_2A'Bi_2I_9$ are 3-4 times higher than those of $Cs_3Bi_2I_9$ within the 100-500 K temperature range, with polar optical phonon scattering identified as the primary mobility-limiting mechanism. Furthermore, the partial substitution of $Cs^+$ with $Ag^+$ or $Cu^+$ enhances the out-of-plane mobility, thereby reducing mobility anisotropy and indicating an increase in electronic dimensionality. Finally, through detailed analysis of longitudinal optical phonon frequencies ($\omega_{LO}$), carrier effective masses ($m^*$), and Fröhlich coupling strengths ($\alpha$), we demonstrate that the enhanced carrier mobility observed in $Cs_2AgBi_2I_9$ and $Cs_2CuBi_2I_9$ is primarily attributed to the small values of $m^*$ and $\alpha$, that is, the small polar mass $\alpha(m^*/m_e)$.

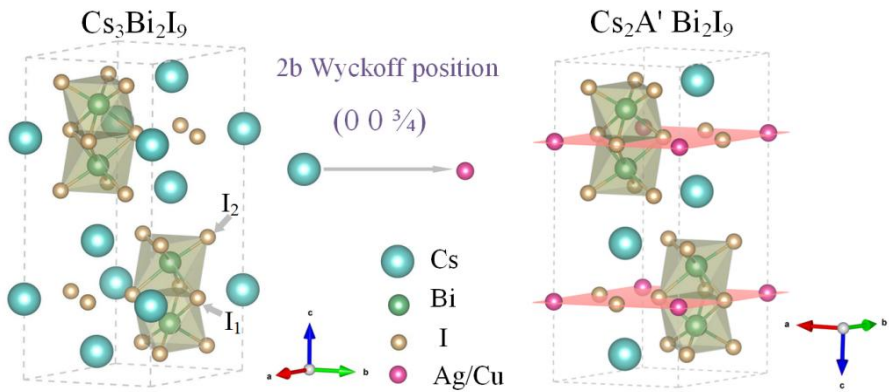

**Figure 1.** Schematic illustration of the transformation from $Cs_3Bi_2I_9$ to $Cs_2AgBi_2I_9$ and $Cs_2CuBi_2I_9$. The left image shows the unit cell structure of $Cs_3Bi_2I_9$, while the right image depicts the unit cell structure of $Cs_2A'Bi_2I_9$ formed by the substitution of $Cs^+$ at the 2b Wyckoff position with $Ag^+$ or $Cu^+$.

According to previous experiments[21], $Cs_3Bi_2I_9$ adopts a hexagonal structure with the $P6_3/mmc$ space group, consisting of $[Bi_2I_9]^{3-}$ dimers, as illustrated in Figure 1. In this structure, $Cs^+$ cations occupy the 4f (2/3, 1/3, z+1/2) and 2b (0, 0, 3/4) Wyckoff positions. By substituting $Ag^+$ for $Cs^+$ at the 2b (0, 0, 3/4) Wyckoff position, the A-site ordered perovskite derivative $Cs_2AgBi_2I_9$ can be obtained, which has been recently synthesized experimentally.[21] Similarly, $Cu^+$ substitution could yield $Cs_2CuBi_2I_9$, although no experimental reports exist for its synthesis to date. Interestingly, $Cu_2AgBiI_6$ has already been experimentally synthesized with $Cu^+$ occupying the A-site.[22] Due to the differences in orbital energies between $Cu^+$ and $Ag^+$, $Cu^+$ substitution in $Cs_3Bi_2I_9$ is expected to further modulate the electronic structure, such as reducing the bandgap.[23]

We performed lattice parameter optimization for $Cs_2AgBi_2I_9$ using various functionals, as shown in Table S1. The results indicate that the PBE functional combined with the D3 van der Waals correction (PBE+D3) provides the best agreement with experimental data.[21] Therefore, this method was selected for structural optimization of all studied systems. The optimized lattice parameters for $Cs_2AgBi_2I_9$ and $Cs_2CuBi_2I_9$ are presented in Table 1, with the values for $Cs_3Bi_2I_9$ also provided for comparison. It can be observed that substituting $Ag^+$ or $Cu^+$ for $Cs^+$ at the 2b Wyckoff position results in a decrease in lattice parameters, volume, and bond lengths. This reduction is attributed to the smaller ionic radii of $Ag^+$ and $Cu^+$ compared to $Cs^+$.[24] Notably, the Ag-I (Cu-I) bond lengths for both interior and exterior iodine atoms within the $[Bi_2I_9]^{3-}$ dimers of $Cs_2A'Bi_2I_9$ (A' = Ag, Cu) are significantly shorter than the Cs-I bond lengths in $Cs_3Bi_2I_9$. This shortening is expected to lead to strong orbital hybridization.

**Table 1.** Structural parameters, volume ($V$), bond lengths, and band gaps ($E_g$) for $Cs_3Bi_2I_9$, $Cs_2AgBi_2I_9$, and $Cs_2CuBi_2I_9$. For comparison, the experimental data from the literature are also provided.

| Compound | $a = b$ (Å) | $c$ (Å) | $V$ (Å$^3$) | Bond length | $E_g$ (eV) |
|---|---|---|---|---|---|

|  |  |  |  | A'-I$_1^a$ | A'-I$_2$ | Bi-I$_1$ | Bi-I$_2$ |  |
|---|---|---|---|---|---|---|---|---|
| Cs$_3$Bi$_2$I$_9$ | 8.52 | 21.52 | 1352.2 | 4.26 | 4.35 | 3.25 | 2.97 | 2.35 |
|  | (8.43$^b$) | (21.34$^b$) | (1314.7$^b$) | (4.22$^b$) | (4.34$^b$) | (3.25$^b$) | (2.94$^b$) | (1.99$^b$) |
| Cs$_2$AgBi$_2$I$_9$ | 8.37 | 21.29 | 1291.3 | 4.19 | 4.19 | 3.21 | 2.97 | 1.28 |
|  | (8.41$^b$) | (21.19$^b$) | (1299.1$^b$) | (4.21$^b$) | (4.31$^b$) | (3.24$^b$) | (2.93$^b$) | (1.72$^b$) |
| Cs$_2$CuBi$_2$I$_9$ | 8.36 | 21.30 | 1288.9 | 4.18 | 4.21 | 3.21 | 2.97 | 0.96 |

$^a$ The positions of I$_1$ and I$_2$ are indicated in Figure 1.   $^b$ Experimental values are obtained from Reference [21].

Figure 2 presents the calculated projected density of states (PDOS), crystal orbital Hamiltonian population (COHP), and atomic orbital hybridization schematics for Cs$_3$Bi$_2$I$_9$, Cs$_2$AgBi$_2$I$_9$, and Cs$_2$CuBi$_2$I$_9$, obtained using the HSE06 method. Note that spin-orbit coupling (SOC) was not included, as COHP analysis does not support SOC. The calculated projected band structures and partial charge densities, including SOC effects, are shown in Figure S1. The specific band gap values reported below are based on the HSE06+SOC method. The results demonstrate that Cs$_3$Bi$_2$I$_9$ is an indirect bandgap semiconductor with a bandgap of 2.35 eV, consistent with previously reported values.[21] The conduction band minimum (CBM) primarily originates from anti-bonding interactions between Bi$^+$ 6p orbitals and I$^-$ 5p orbitals, while the valence band maximum (VBM) mainly arises from anti-bonding interactions between Bi$^+$ 6s orbitals and I$^-$ 5p orbitals. Notably, the bandgap of Cs$_2$AgBi$_2$I$_9$, formed by partially substituting Cs$^+$ with Ag$^+$, is significantly reduced to 1.28 eV. The PDOS and COHP analysis of Cs$_2$AgBi$_2$I$_9$ reveals that the reduction in the bandgap is due to the formation of intermediate energy bands, which constitute the new CBM and are primarily composed of contributions from Ag$^+$ 5s orbitals. It's noteworthy that the orbital contribution of Ag$^+$ in A-site ordered Cs$_2$AgBi$_2$I$_9$ is distinctly different from that in B-site ordered Cs$_2$AgBiBr$_6$, where Ag$^+$ 4d orbitals contribute to the VBM. Similarly, in Cs$_2$CuBi$_2$I$_9$, intermediate conduction bands are also formed, primarily from contributions of Cu$^+$ 4s orbitals (Figures 2e, 2f). Due to the lower energy level of Cu⁺ 4s compared to Ag⁺ 5s, the introduction of Cu⁺ into $Cs_3Bi_2I_9$ further reduces the bandgap to 0.96 eV. In addition to significantly reducing the bandgap, partially substituting $Ag^+/Cu^+$ for $Cs^+$ also noticeably improves the band edge dispersion and reduces the carrier effective mass, which will be discussed in detail in a later section.

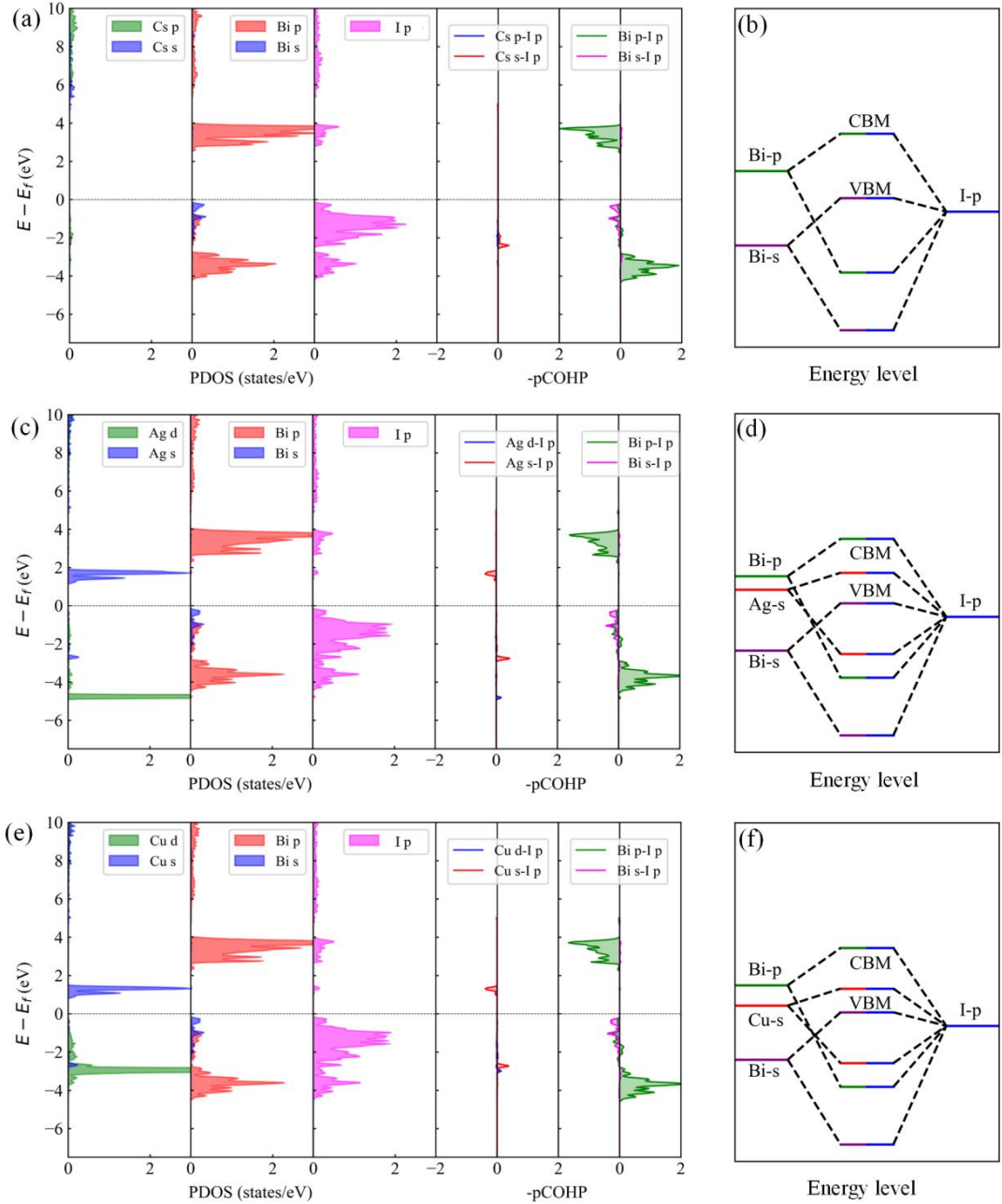

**Figure 2.** Projected density of states (PDOS), crystal orbital Hamiltonian population (COHP), and schematic atomic orbitals and coupling models for (a) $Cs_3Bi_2I_9$, (c) $Cs_2AgBi_2I_9$, and (e) $Cs_2CuBi_2I_9$.

To better understand how substituting $Ag^+/Cu^+$ for $Cs^+$ affects the nature of charge transport, first-principles carrier mobility calculations were performed using the AMSET software.[25] Both n-type and p-type doping were investigated, with calculations including scattering from ionized impurities (IMP), acoustic phonons (ADP), and polar optical phonons (POP). Piezoelectric scattering was excluded due to the centrosymmetric crystal structures of the three studied systems. Additional details on the mobility calculations and the underlying principles of AMSET can be found in the computational methods section and the relevant literature. [25]

Figures 3a,3b show the calculated average mobilities of electrons and holes in $Cs_3Bi_2I_9$ and $Cs_2A'Bi_2I_9$ (A' = Ag, Cu) as a function of bulk defect concentration at 300 K. The mobilities for all three compounds remain relatively constant within the defect concentration range of $10^{12}$ cm$^{-3}$ to $10^{16}$ cm$^{-3}$. However, above $10^{16}$ cm$^{-3}$, a significant decrease in mobility is observed for all three compounds. Notably, the average mobility of $Cs_2AgBi_2I_9$ and $Cs_2CuBi_2I_9$ is approximately 3-4 times higher than that of $Cs_3Bi_2I_9$ for both electrons and holes. For example, at moderate defect concentrations ($10^{16}$ cm$^{-3}$), the averaged electrons/holes mobilities at room temperature for $Cs_3Bi_2I_9$, $Cs_2AgBi_2I_9$, and $Cs_2CuBi_2I_9$ are 7.20/4.51 cm$^2$/Vs, 29.51/11.52 cm$^2$/Vs, and 27.06/12.50 cm$^2$/Vs respectively. It is noteworthy that, compared to introducing $Ag^+$ at the A-site of $Cs_3Bi_2I_9$, substituting with $Cu^+$ further enhances hole mobility. The calculated mobilities for $Cs_2AgBi_2I_9$ and $Cs_2CuBi_2I_9$ are comparable to the previously reported theoretical values at 300 K for $Sb_2S_3$ (15~40 cm²/Vs) and $MAPbI_3$ (30~80 cm²/Vs).[26,27] Furthermore, electron mobility is significantly higher than hole mobility in all three materials, suggesting doping asymmetry and indicating that n-type doping could be beneficial for carrier collection in optoelectronic devices. This phenomenon is also observed in other optoelectronic semiconductors. Further analysis of the individual contributions from different scattering mechanisms reveals that, across the range of low to high defect concentrations and varying temperatures, polar optical phonon scattering remains the dominant mechanism limiting carrier mobilities (Figure 3c and Figure S3-S5). This finding is consistent with reports on other halide perovskites.[25,27]

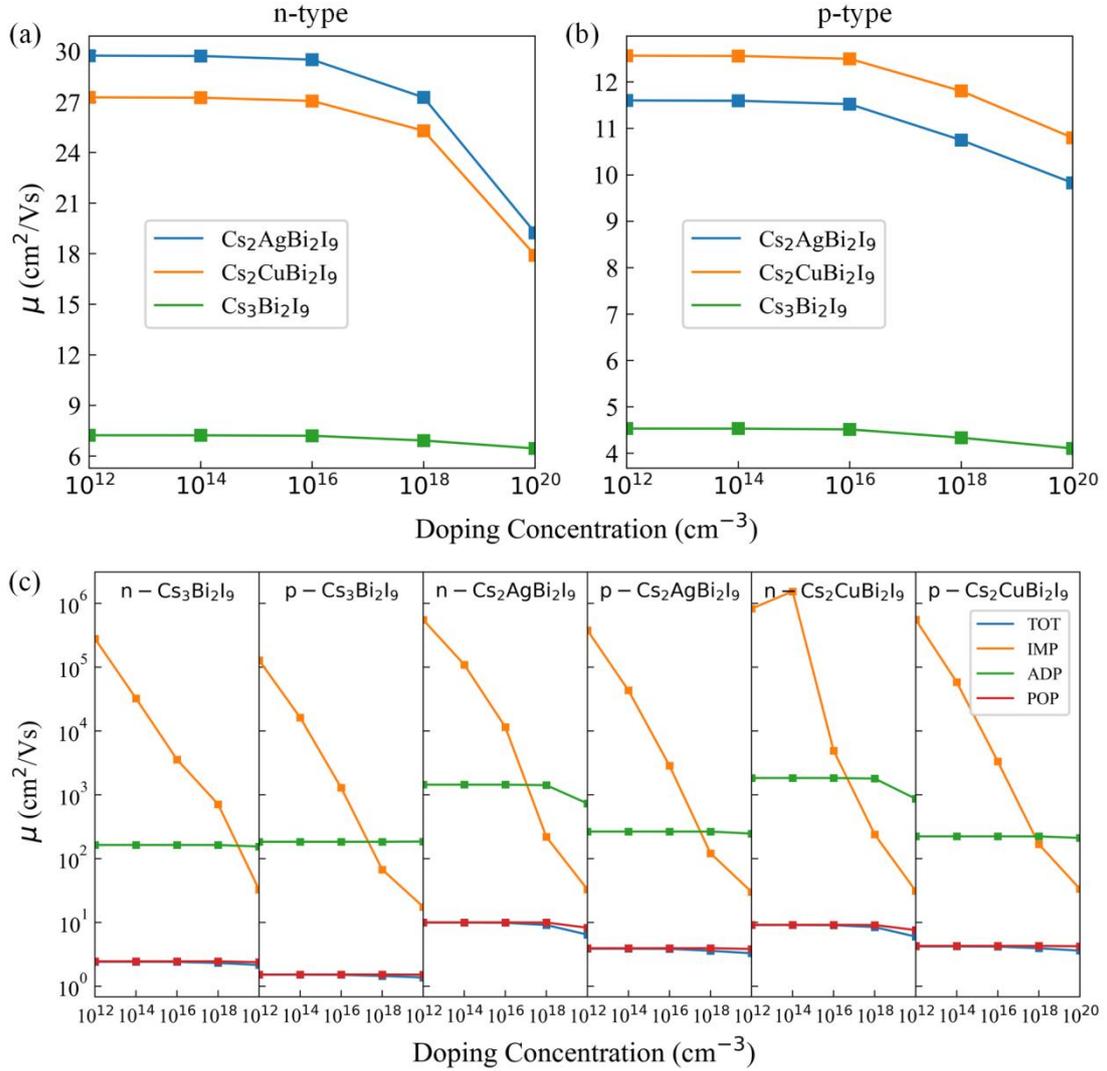

**Figure 3.** The total carrier mobility as a function of doping concentration at 300 K: (a) n-type, (b) p-type, and (c) contributions from different scattering mechanisms as a function of doping concentration.

The anisotropy of carrier mobility was also investigated. Figure 4 illustrates the calculated out-of-plane (*z*-axis) and in-plane (*xy*-plane) mobilities of electrons and holes in $Cs_3Bi_2I_9$ and $Cs_2A'Bi_2I_9$ (A' = Ag, Cu) as a function of temperature. As the temperature increases from 100 K to 500 K, mobility decreases in each direction for all three compounds. Notably, $Cs_3Bi_2I_9$ exhibits significant anisotropy in mobility between the in-plane and out-of-plane directions for both electrons and holes, especially at lower temperatures. Interestingly, the partial substitution of $Cs^+$ ions with Ag$^+$ or Cu$^+$ in Cs$_3$Bi$_2$I$_9$ results in a substantial enhancement of the out-of-plane mobility, particularly at low temperatures, accompanied by a reduction in mobility anisotropy. For examples, the electron (hole) mobilities along the *x*/*y*/*z* directions (10$^{16}$ cm$^{-3}$ and 100 K) for Cs$_3$Bi$_2$I$_9$, Cs$_2$AgBi$_2$I$_9$ and Cs$_2$CuBi$_2$I$_9$ are 13.56/12.74/1.03 (3.88/3.58/9.38) cm$^2$/Vs, 50.85/53.38/43.16 (21.02/20.22/17.98) cm$^2$/Vs, and 40.13/41.97/35.73 (19.22/18.66/16.43) cm$^2$/Vs, respectively. This suggests that Cs$_2$A'Bi$_2$I$_9$ (A' = Ag, Cu) exhibits an increased electronic dimensionality compared to Cs$_3$Bi$_2$I$_9$, which is attributed to reduced bond lengths and enhanced orbital coupling.

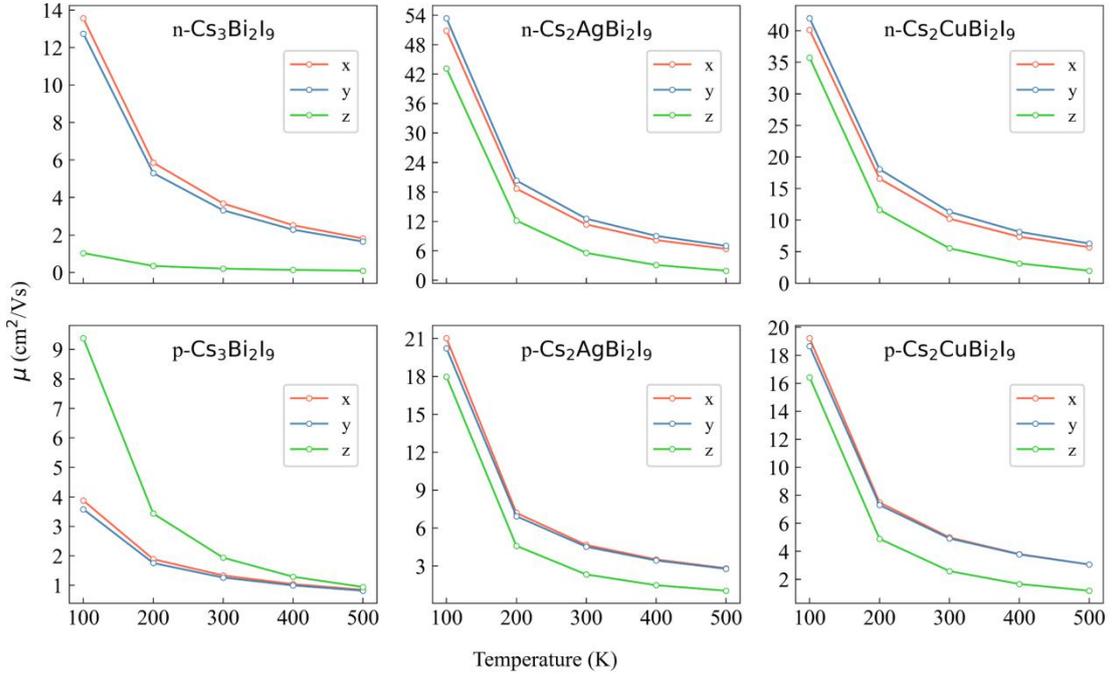

**Figure 4.** Temperature-dependent carrier mobility of Cs$_3$Bi$_2$I$_9$, Cs$_2$AgBi$_2$I$_9$, and Cs$_2$CuBi$_2$I$_9$ at a doping concentration of 10$^{16}$ cm$^{-3}$. The panels display the mobility profiles along *x*, *y*, and *z* crystallographic directions for both electron (n-type) and hole (p-type) carriers across the temperature range from 100 K to 500 K.

In previous discussions, we established that polar optical phonon scattering is the dominant mechanism limiting carrier mobility in Cs$_3$Bi$_2$I$_9$ and Cs$_2$A'Bi$_2$I$_9$ (A' = Ag, Cu). Giustino et al. previously proposed a universal scaling law to understand the mobility trends in halide perovskites.[28] Their simplified conceptual model suggests that mobility is determined exclusively by two dimensionless parameters: the polar mass $\alpha(m^*/m_e)$ and the reduced frequency $\hbar\omega_{LO}/k_BT$.[28] Here, $\alpha$ denotes the Fröhlich coupling strength, $m^*/m_e$ is the effective mass relative to the electron mass, and $\omega_{LO}$ is the frequency of the longitudinal optical (LO) phonon. High mobilities can be achieved when the polar mass is small and the phonon frequency is large.

To investigate why the mobilities of $Cs_2A'Bi_2I_9$ (A' = Ag, Cu) are 3-4 times larger than those of $Cs_3Bi_2I_9$ at room temperature, we calculated the corresponding physical parameters, including $m^*$, $\omega_{LO}$, and $\alpha$, as shown in Table 2. The parameter $\alpha$ was determined using the following formula,[29]

$$\alpha = \frac{1}{4\pi\varepsilon_0}\frac{1}{2}\left(\frac{1}{\varepsilon_\infty} - \frac{1}{\varepsilon_{static}}\right)\frac{e^2}{\hbar\omega_{LO}}\left(\frac{2m^*\omega_{LO}}{\hbar}\right)^{1/2} \quad (1)$$

where $\varepsilon_\infty$ is the optical dielectric constant and $\varepsilon_{static}$ is the static dielectric constant. As shown in Table 2, the calculated electron/hole Fröhlich coupling strengths for $Cs_3Bi_2I_9$, $Cs_2AgBi_2I_9$, and $Cs_2CuBi_2I_9$ are 12.67/8.14, 2.16/3.85, and 1.68/2.3, respectively. This indicates that the introduction of $Ag^+$ and $Cu^+$ at the A site significantly weakens the Fröhlich coupling strength, which is consistent with the experimentally observed weak electron-phonon coupling. From an electronic structure perspective, partial substitution of $Cs^+$ with $Ag^+$ or $Cu^+$ is expected to enhance the orbital coupling between $Ag^+/Cu^+$ s and $I^-$ p and increase band dispersion due to the reduced bond lengths. Consequently, the carrier effective masses ($m^*$) in $Cs_2AgBi_2I_9$ and $Cs_2CuBi_2I_9$ are significantly smaller than those in $Cs_3Bi_2I_9$. Furthermore, the $\omega_{LO}$ values for $Cs_3Bi_2I_9$, $Cs_2AgBi_2I_9$ and $Cs_2CuBi_2I_9$ are 2.04, 2.34, and 2.34, respectively. Notably, the $\omega_{LO}$ values of $Cs_2A'Bi_2I_9$ (A' = Ag, Cu) show only a slight increase relative to $Cs_3Bi_2I_9$. Therefore, the enhanced carrier mobility observed in $Cs_2AgBi_2I_9$ and $Cs_2CuBi_2I_9$ is primarily attributed to the small polar mass $\alpha(m^*/m_e)$.

**Table 2** Calculated dielectric constants ($\varepsilon$), effective phonon frequency ($\omega_{LO}$, in THz), carrier effective masses ($m^*$, in $m_e$), and Fröhlich parameter ($\alpha$) for the in-plane direction in $Cs_3Bi_2I_9$, $Cs_2AgBi_2I_9$, and $Cs_2CuBi_2I_9$. Note that due to the relatively flat

bands along the out-of-plane direction, the calculated carrier effective masses are subject to significant errors in that direction. Therefore, only the in-plane results for the three studied compounds are presented here.

|  | $\varepsilon_\infty$ | $\varepsilon_{static}$ | $\omega_{LO}$ | $m^*$ | | $\alpha$ | |
|---|---|---|---|---|---|---|---|
|  |  |  |  | $e^-$ | $h^+$ | $e^-$ | $h^+$ |
| $Cs_3Bi_2I_9$ | 4.39 | 9.85 | 2.04 | 2.58 | 6.25 | 8.14 | 12.67 |
| $Cs_2AgBi_2I_9$ | 5.67 | 10.54 | 2.34 | 0.5 | 1.59 | 2.16 | 3.85 |
| $Cs_2CuBi_2I_9$ | 6.48 | 10.62 | 2.34 | 0.53 | 1.04 | 1.68 | 2.3 |

In summary, we thoroughly investigated the electronic properties and carrier transport properties in A-site ordered $Cs_2A'Bi_2I_9$ (A' = Ag, Cu) using first-principles calculations combined with Boltzmann transport calculations. Our results reveal that the introduction of $Ag^+$ or $Cu^+$ cations with smaller ionic radii at the A-site of $Cs_3Bi_2I_9$ significantly reduces the A'-I bond length, leading to enhanced orbital hybridization and band edge dispersion. Additionally, the $Ag^+$ 5s or $Cu^+$ 4s orbitals form an intermediate band that constitutes a new lower conduction band. Therefore, compared to the band gap of $Cs_3Bi_2I_9$ (2.35 eV), $Cs_2AgBi_2I_9$ and $Cs_2CuBi_2I_9$ exhibit significantly reduced band gaps, which are 1.28 eV and 0.96 eV, respectively.

The transport properties calculations reveal that the carrier mobilities of A-site ordered $Cs_2A'Bi_2I_9$ are 3-4 times higher than those of $Cs_3Bi_2I_9$ over the temperature range of 100-500 K. For example, at a defect concentration of $10^{16}$ cm$^{-3}$, the electron/hole mobilities at room temperature are 7.20/4.51 cm$^2$/Vs for $Cs_3Bi_2I_9$, 29.51/11.52 cm$^2$/Vs for $Cs_2AgBi_2I_9$, and 27.06/12.50 cm$^2$/Vs for $Cs_2CuBi_2I_9$. Importantly, polar optical phonon scattering is the primary mechanism limiting carrier mobility in these three compounds. Moreover, the partial substitution of $Ag^+$ or $Cu^+$ for $Cs^+$ significantly increases the out-of-plane carrier mobility (i.e., along the z direction), leading to reduced mobility anisotropy and indicating increased electronic dimensionality. This improvement is mainly due to shorter bond lengths and stronger orbital coupling between $Ag^+/Cu^+$ s orbitals and I p orbitals in $Cs_2A'Bi_2I_9$.

Finally, based on a simplified conceptual model, through detailed analysis of longitudinal optical phonon frequencies ($\omega_{LO}$), carrier effective masses ($m^*$), and Fröhlich coupling strengths ($\alpha$), we demonstrate that the enhanced carrier mobility observed in $Cs_2AgBi_2I_9$ and $Cs_2CuBi_2I_9$ is primarily attributed to the small values of $m^*$ and $\alpha$, that is, the small polar mass $\alpha(m^*/m_e)$. Our study offers significant insights into the transport properties of Bi-based perovskite derivatives, facilitating their advancement for use in optoelectronic applications.

First-principles calculations were performed using density functional theory (DFT) with the projector augmented wave (PAW) method as implemented in the Vienna ab initio Simulation Package (VASP).[30] Based on comparisons with experimental structural data, the Perdew-Burke-Ernzerhof (PBE) exchange-correlation functional was selected, and DFT-D3 dispersion corrections were included to improve the accuracy of lattice constant predictions.[31,32] Geometry optimizations were conducted with a plane-wave energy cutoff of 520 eV and a Γ-centered 5×5×2 $k$-point mesh.[33] The convergence thresholds of the total energy and the Hellmann-Feynman force are $10^{-7}$ eV and 0.001 eV/Å, respectively. Spin-orbit coupling (SOC) was included in the band structure calculations. Since PBE tends to underestimate the band gap, the hybrid functional HSE06 was employed for more accurate band gap predictions.[34]

The electrical transport properties of the selected materials were computed using the AMSET code.[25] The required input parameters, including high-frequency and static dielectric constants, elastic constants, and polar optical phonon frequencies, were all determined through first-principles calculations (detailed values can be found in the Supporting Information). The high-frequency and static dielectric constants, as well as the effective polar optical phonon frequency, were estimated using Density Functional Perturbation Theory (DFPT).[35] The elastic constants were calculated using the energy-strain method with the PBE functional. Special attention was given to the spin-orbit coupling effect when calculating the deformation potential and energy eigenvalues. To balance computational cost and accuracy, the *k*-point meshes used for calculating transport properties were tested (as shown in Figure S2), and a 55×55×19 *k*-point mesh was ultimately chosen for all calculations.

Please refer to the supplementary materials for detailed information on the projected band structures, the partial charge density of the VBM (left) and CBM (right), as well as the details of the carrier mobility calculations.

This work at Beijing Institute of Technology was supported by the National Key Research and Development Program of China (2021YFA1400300), the National Natural Science Foundation of China (Grant No. 12172047), and Beijing National Laboratory for Condensed Matter Physics (2023BNLCMPKF003). G. T. was supported by Beijing Institute of Technology Research Fund Program for Young Scholars (Grant No. XSQD-202222008) and Guangdong Key Laboratory of Electronic Functional Materials and Devices Open Fund (EFMD2023004M).